\documentclass[journal=jctcce,manuscript=article,articletitle=true,layout=onecolumn]{achemso}

\usepackage{amsmath}
\usepackage{amssymb}
\usepackage{bm}
\usepackage{latexsym}
\usepackage{graphicx}
\usepackage{epsfig,epsf,rotating}
\usepackage{indentfirst}
\usepackage{epstopdf}
\usepackage{xcolor}
\usepackage{multirow}
\usepackage{tabularx}
\usepackage{ctable}

\usepackage{upgreek}
\newcolumntype{C}{>{\centering\arraybackslash}X}
\usepackage{nicefrac}

\title{Consequences of the failure of equipartition for the $p-V$ behavior of liquid water and the hydration free energy components of a small protein\footnote{
Notice:  This manuscript has been authored by UT-Battelle, LLC, under contract DE-AC05-00OR22725 with the US Department of Energy (DOE). The US government retains and the publisher, by accepting the article for publication, acknowledges that the US government retains a nonexclusive, paid-up, irrevocable, worldwide license to publish or reproduce the published form of this manuscript, or allow others to do so, for US government purposes. DOE will provide public access to these results of federally sponsored research in accordance with the DOE Public Access Plan (https://www.energy.gov/doe-public-access-plan}}
\author{Dilipkumar N. Asthagiri}
\affiliation{Oak Ridge National Laboratory, One Bethel Valley Road, Oak Ridge, TN 37830}
\email{asthagiridn@ornl.gov}
\author{Arjun Valiya Parambathu}
\affiliation{Chemical and Biomolecular Engineering, University of Delaware, Newark, DE 19700}
\author{Thomas L. Beck}
\affiliation{Oak Ridge National Laboratory, One Bethel Valley Road, Oak Ridge, TN 37830}

\begin{document}
\clearpage
\begin{abstract}
Earlier we showed that in the molecular dynamics simulation of  a rigid model of water it is necessary to use an integration time-step $\delta t \leq 0.5$~fs to ensure equipartition between translational and rotational modes. Here we extend that study in the $NVT$ ensemble to $NpT$ conditions and to an aqueous protein. We study neat liquid water with the rigid, SPC/E model and the protein BBA (PDB ID: 1FME) solvated in the rigid, TIP3P model. We examine integration time-steps ranging from $0.5$~fs to $4.0$~fs for various thermostat plus barostat combinations.  We find  that a small $\delta t$ is necessary to ensure consistent prediction of the simulation volume.  Hydrogen mass repartitioning alleviates the problem somewhat, but is ineffective for the typical time-step used with this approach. The compressibility, a measure of volume fluctuations, and the dielectric constant, a measure of dipole moment fluctuations, are also seen to be sensitive to $\delta t$.  Using the mean volume estimated from the $NpT$ simulation, we examine the electrostatic and van der Waals contribution to the hydration free energy of the protein in the $NVT$ ensemble.  These contributions are also sensitive to $\delta t$. In going from $\delta t = 2$~fs to $\delta t = 0.5$~fs, the change in the net electrostatic plus van der Waals contribution to the hydration of BBA is already in excess of the folding free energy reported for this protein. 
\begin{tocentry}
\center{\includegraphics[scale=0.69]{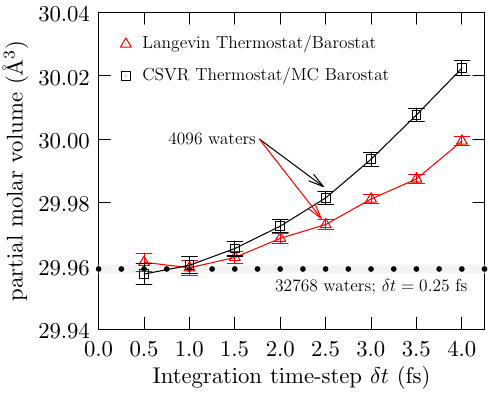}}
\end{tocentry}

 \end{abstract}

\clearpage
\section{Introduction}

Liquid water is the pre-eminent solvent for biological, geological, and chemical processes. Consistent with its pervasive role, it has been widely studied both experimentally and theoretically.  In the theoretical and simulation context, modeling the intermolecular interactions and simulating liquid water has occupied a central place in the overall enterprise of computer simulation of materials. 

Molecular dynamics simulations of water have a rich history.  The seminal work by 
Rahman and Stillinger\cite{rahman:jcp71,stillinger:jcp74} over half a century ago stands as a towering initial attempt to model the structure and dynamics of water. They described water as a rigid object, treating the translational motion using cartesian coordinates and the rotational motion using Euler angles. Through careful analysis, they settled on a time-step $\delta t = 0.4$~fs for integrating the equations of motion. 
Some years later, Ryckaert, Ciccotti, and Berendsen introduced the SHAKE algorithm\cite{md:shake} that enabled treating molecules such as water as a rigid object within a cartesian coordinate system.  SHAKE had the defect of not accounting for zero relative velocity of atoms connected by a rigid bond. This was fixed by the RATTLE algorithm\cite{andersen:rattle83}.  The subsequent development of the analytical SETTLE algorithm\cite{kollman:settle92} that obeys both the rigidity and velocity constraints for 3-site water molecules was another important innovation with significant impact in bio-molecular simulations. 

In developing SHAKE, the authors  introduced a seemingly innocuous assumption, namely ``fast internal vibrations are usually decoupled from rotational and translational motions''\cite{md:shake}. This idea --- an early form of coarse-graining ---  has become central to modeling rigid-water molecular dynamics with time-steps that are considerably larger than the value used by Rahman and Stillinger. Most often, a value of 2~fs has been used in many studies over the last 4+ decades.  It must then come as a surprise that recently we found that this assumption of decoupling of vibrations from rotations is not valid for water\cite{asthagiri:jctc2024a}:  the rotational relaxation occurs on time-scales comparable to bond vibration and angle bending modes of water. A consequence of this failure is that with a long time-step, such as the commonly used 2~fs, equipartition between translation and rotation is broken.  Ultimately, this 
breakdown is a consequence of not capturing the translational and rotational relaxation with fidelity, as is required for obeying the fluctuation-dissipation relation for the respective modes \cite{pathria}. Our finding reconfirmed the breakdown of equipartition for bulk water noted in earlier works by Davidchack\cite{davidchack:jcp10} and Silveira and Abreu\cite{abreu:jcp17,abreu:jcp19}.

 For time-steps that lead to a breakdown of equipartition, we had found that the center of mass motion of water is at a higher temperature than the rotational motion about the center of mass. This suggested that for a constant volume simulation, the pressure must be higher for a larger $\delta t$. Indeed, Davidchack had found exactly that behavior. We reasoned that under $NpT$ conditions,  
 the volume must be higher for a higher $\delta t$. We validate this hypothesis in the present paper for both bulk water and for the designed protein with a $\beta\beta\alpha$ fold\cite{mayo:sc1997} , hereafter termed protein BBA, solvated in water. For the bulk water simulation, the difference in volume between simulations with $\delta t = 0.5$~fs and $\delta t = 2.0$~fs can easily exceed typical volume changes in protein conformational change and protein unfolding. We also test the idea of hydrogen mass repartitioning \cite{hopkins:hmr2015}, and find that while this helps somewhat, equipartition is still violated for the typical time-steps used with this approach. 

We further reasoned that the isothermal compressibility and the static dielectric constant, properties that derive from the fluctuation of a related extensive quantity, should also depend on $\delta t$.  The calculation of these properties from the appropriate fluctuation relationship requires rather long simulations for adequate convergence. (For example, see Ref.\ \citenum{fuentes:dielectric14}.) Nevertheless, the dependence of the isothermal compressibility on $\delta t$ can still be inferred from the simulations. For the dielectric constant, we find it easier to calculate the electrostatic contribution to the free energy of hydration. For the protein BBA in water, both the electrostatic and van der Waals (vdW) contribution to hydration 
depend on $\delta t$. Importantly, in going from $\delta t = 2.0$~fs and $\delta t = 0.5$~fs, the change in the sum of the electrostatic and vdW contributions, an approximation to the net hydrophilic contribution in hydration, is comparable to the free energy of folding reported for BBA \cite{shea:bba2010,deshaw:sc2011}.  



 \section{Methodology} 
 
\subsection{Bulk water} 

We studied the SPC/E\cite{spce} water model using both NAMD\cite{namd,namd:2020} and GPU-accelerated Tinker\cite{tinker8,tinker9} codes. The system comprised 4096 water molecules. Throughout, the equations of motion were integrated using the Velocity Verlet algorithm. 

In NAMD,  the Lennard-Jones and real-space electrostatic interactions were cutoff at 9~{\AA}. Long-range, analytical LJ corrections were applied.  The electrostatic interactions were calculated using particle mesh Ewald with a grid spacing of 1~{\AA}. The relative Ewald energy tolerance at the real-space cutoff was $10^{-7}$, tighter than the default in NAMD.  The system was first equilibrated for 6~ns under $NVT$ conditions at a mass density of 1.014 gm/cc ($\approx1.5$\% higher than the value noted for SPC/E\cite{sanz:prl04}).  We used the canonical stochastic velocity rescaling thermostat (CSVR)\cite{svr:jcp07} to maintain the system at 298.15~K and the time-step in this phase was 0.5~fs. This initial structure was used in all subsequent studies.

The initial structure obtained above was next equilibrated under $NpT$ conditions for $8\times 10^6$ steps followed by a production phase of $20\times 10^6$ steps. We simulated using time-steps ranging from 0.5~fs to 4.0~fs in 0.5~fs interval. The geometry of the water molecule was maintained using the default SETTLE algorithm. Simulation data (energies, volumes, etc.) were archived every 500 steps for further analysis. 

Within Tinker, we used the default Ewald cutoff of 7~{\AA} and the default Lennard-Jones cutoff of 9~{\AA}. Long-range LJ corrections were applied. The system was simulated for a total of $28\times 10^6$ steps as above, but data logging frequency changed with step size, being approximately every 1~ps. (We say approximately since some time-steps do not evenly divide 1~ps.) The geometry of water was maintained using the RATTLE algorithm. For simulations with Tinker, we also studied the effect of mass repartitioning\cite{hopkins:hmr2015} --- the mass of the hydrogen atom in water was increased to 3.024 amu and the mass of oxygen appropriately reduced such that the mass of a water molecule remained at 18.0154 amu. 

With NAMD, we experimented with the following thermostat-barostat combinations: Langevin thermostat/barostat\cite{feller:jcp95}, CSVR thermostat/Langevin barostat, and CSVR thermostat/Monte Carlo barostat \cite{chow:mcbarostat1995,aqvist:mcbarostat2004}. With Tinker, we used the Monte Carlo barostat and the CSVR thermostat. For the volume sampling frequency in the Monte Carlo barostat, we used the default values in the respective codes: 50 steps in NAMD and 25 steps in Tinker.  

Within Tinker, for a limited set of runs, we experimented using the Beeman algorithm\cite{beeman:1976} to learn if an improved estimate of on-step velocity affected the overall conclusions. These simulations were  
performed exclusively on CPUs. 

Throughout, we use the Friedberg-Cameron approach \cite{friedberg:1970,allen:error} to obtain statistical uncertainties for quantities such as the volume, the potential energy of the 
system, or the binding energy between a solute and the solvent (see below). 

\subsubsection{Bulk reference}

To provide a separate estimate of convergence of volume, we studied a larger 32768 water system with a time-step of $\delta t = 0.25$~fs. 
This system was obtained by replicating the 4096 water system twice in the $x$, $y$, and $z$ directions, respectively.  The simulation box length was set to 100 {\AA};  the bulk density
was $\approx2$\% less than the converged value we found with the $\delta t = 0.5$~fs simulations.  We equilibrated this 
system in the $NVT$ ensemble for $6\times 10^6$ time-steps using the CSVR thermostat.  The equilibrated configuration was then used to launch four separate $NpT$ ensemble simulations using the CSVR thermostat and Monte Carlo barostat.  The volume sampling frequency for the barostat was 80, 120, 160, and 200 time-steps, respectively, for the four separate runs. The $NpT$ simulations were equilibrated for $6.25\times 10^6$ steps and data collected over an additional $6.25\times 10^6$ steps. In reporting the data for this larger system, the mean from the four separate runs are averaged and the standard error of the mean combined using variance propagation rules.

\subsection{Aqueous BBA} 

BBA (PDB ID: 1FME) is a 28 residue designed protein that adopts a $\beta\beta\alpha$ fold. This is a marginally stable protein derived 
from a parent zinc-finger template\cite{mayo:sc1997} sans the zinc ion. The first model from the PDB data file was taken and solvated in 6561 TIP3P\cite{tip32} water molecules. The N- and C-termini were modeled in the ammonium and carboxylate forms, respectively.  At pH 7.0, the protein has a net charge of $+4e$, where $e$ is the elementary charge; the net charge of the protein was compensated by adding 4 chloride (Cl$^{-}$) ions to the system.  The protein was modeled using the CHARMM36m\cite{charmm,cmap2,charmm36,charmm36m:2016} forcefield (including CMAP corrections). CHARMM-modified parameters were used for TIP3P\cite{tip3mod}.  The initial structure was built using the {\sc psfgen} tool\cite{psfgen} and the chloride ions were added using the {\sc autoionize} tool within {\sc VMD} \cite{hump:1996}. 

In the first set of simulations, no structural constraints were placed on the protein. The initial system was  equilibrated under $NpT$ conditions using $\delta t = 0.5$~fs for 4~ns ($8\times 10^6$ steps).  The Lennard-Jones forces were smoothly switched to zero from 9.43 {\AA} to 10.43 {\AA}. The particle mesh Ewald method was used for long-range electrostatic interactions, and as above, a tighter tolerance was used for Ewald summations. The bond between a hydrogen and the parent heavy atom was made rigid using the {\sc rigidbonds all} flag in NAMD.  In this phase of equilibration, we used the Langevin thermostat and barostat.  

The configuration from the end-point of the equilibration run was used to launch simulations at time-steps from 0.5~fs to 3.5~fs in intervals of 0.5~fs. For these studies we experimented with the following thermostat and barostat combinations: Langevin thermostat/barostat and CSVR thermostat/Monte Carlo barostat. For each $\delta t$, the system was equilibrated over $12\times 10^6$ steps and data collected over a subsequent $20\times 10^6$ steps, with data logged every 500 steps. The simulation trajectory was archived every 1000 steps for further analysis. 

\subsubsection{Hydration free energy components of BBA}

For calculating the hydrophilic contributions to the hydration free energy of BBA, we made two important changes. First, we fixed the protein conformation. Second, we removed the Cl$^{-}$ ions. (Within the Ewald formulation, the uniform compensating background charge ensures the electroneutrality of the system \cite{Hummer:ions1996,Hummer:ions1998}.) 
These changes were made to ensure that protein conformational fluctuations or variation in the binding of Cl$^{-}$ ions to the protein do not obfuscate the role of $\delta t$. The initial protein conformation was obtained by scanning the $NpT$ simulation (using the CSVR thermostat/MC barostat and $\delta t = 0.5$~fs) for the conformer with the least deviation from the reference 1FME conformer.  The RMS deviation of the chosen structure relative to the original 1FME structure was 1.55 {\AA}. 

Since we fix the conformation of the protein, we cannot use GPU-resident calculations using NAMD. (For the same reason, we cannot use the Monte Carlo barostat, as this is only available in the GPU-resident mode.) Thus, we used the CSVR thermostat and Langevin barostat to equilibrate the volume. Once the volume stabilized, we removed the barostat as well. The hydration free energy calculations were then performed in the $NVT$ ensemble. 

The electrostatic contribution to the free energy, $\mu^{({\rm ex})}_{elec}$, was obtained by a thermodynamic integration procedure using a three point Gauss-Legendre quadrature \cite{Hummer:jcp96,Weber:jctc12}, with protein charges scaled by $\lambda = \{0.5, 0.5\pm\sqrt{3/20}\}$.   Specifically, 
\begin{equation}
\mu^{({\rm ex})}_{elec} = Q \int_0^1 \langle \phi \rangle_{\lambda} d\lambda \approx \frac{Q}{2} \sum_{\lambda} w_\lambda \langle \phi \rangle_{\lambda} \, 
\end{equation} 
where $\langle \phi \rangle_{\lambda}$ is the electrostatic contribution to the interaction energy between the protein and the solvent with configurations sampled from the ensemble with charges scaled by $\lambda$; $w_\lambda$ is the weight associated with the sampling point $\lambda$; and $Q = +4e$ is the net charge of the protein. (N.B. $\lambda = 0.5$
gives the linear-response estimate of $\mu^{(\rm{ex})}_{elec}$.)  To complete the calculation, it is necessary to consider (Wigner) self-interaction \cite{Hummer:ions1996,Hummer:ions1998} and finite-size\cite{Hummer:ionsize1997} corrections. These scale with the length, $L$, of the cubic simulation box as $1/L$ and $(R/L)^2 \cdot 1/L$, respectively, where $R$ is the nominal radius of the protein. Although the volume changes with $\delta t$, as discussed below, the impact on the change in the Wigner self interaction contribution proves to be small, especially between $\delta t = 0.5$~fs and $\delta t = 2.0$~fs. Further, since the protein occupies a small volume of the simulation cell, we ignore finite-size corrections as well. 

At each $\lambda$ point, the system was equilibrated for $7.5\times 10^6$ steps and configurations archived every 500 steps in the subsequent production run of $7.5\times 10^6$ steps. The {\sc pair interaction} approach in NAMD was used to calculate $\langle \phi \rangle_\lambda$.  

The van der Waals (or nonpolar) contribution, $\mu^{(\rm{ex})}_{vdW}$,  to the free energy of hydration can be calculated from the quasichemical organization of the potential distribution theorem\cite{lrp:apc02,lrp:book} for a solute with all partial charges set to zero. Specifically\cite{asthagiri:jacs07}, 
\begin{eqnarray}
\mu^{(\rm{ex})}_{vdW} = \mu^{(\rm{ex})}_{HC} + k_{\rm B}T \ln p(n=0) + k_{\rm B}T \ln \int P(\varepsilon|n=0) e^{\beta \varepsilon} d\varepsilon \, ,
\end{eqnarray}
where $ \mu^{(\rm{ex})}_{HC}$ is the hard-core (or packing) contribution to the hydration from a solute that simply excludes the solvent from a volume comprising the solute plus a defined inner shell; $P(\varepsilon|n=0)$ is the probability distribution of the binding energy of the solute with the solvent, subject to the inner-shell being bereft ($n=0$) of solvent; $p(n=0)$ is the associated marginal distribution; and $\beta = 1/k_{\rm B}T$ is the reciprocal temperature in energy units.  
A rigorous calculation of the terms in the above equation has been presented in the past \cite{asthagiri:jacs07,asthagiri:jcp2008,tomar:bj2013,tomar:jpcb16,tomar:jpcl20}, but such calculations are demanding and require supplying external forces to NAMD using the Tcl interface. In Ref.\ \citenum{asthagiri:jctc2024a} we have already established that for a small cavity, $ \mu^{(\rm{ex})}_{HC}$ is sensitive to $\delta t$.  Since our interest here is mainly to detect the role of $\delta t$ in the protein-solvent interaction energy, we adopted the following procedure that will be accessible to most users of simulation codes. We completely ignored the excluded volume contribution and considered the molecular envelope as the inner shell, thus $p(n=0) =1$.  We then computed the binding energy distribution, $P(\varepsilon)$, of the protein with the solvent. If $P(\varepsilon)$ is Gaussian with mean $ \langle \varepsilon \rangle$ and variance $\sigma^2$, and subject to the aforementioned simplifications, the vdW contribution, $\mu^{(\rm{ex})}_{vdW}$, is given by\cite{asthagiri:jacs07,lrp:book} \begin{eqnarray}
\mu^{(\rm{ex})}_{vdW} = \langle \varepsilon \rangle + \beta \frac{\sigma^2}{2} \,  . 
\end{eqnarray}
We find that the difference in $\mu^{({\rm ex})}_{vdW}$ between adjacent $\delta t$ values is insensitive to whether or not we include the fluctuation contribution ($\beta \sigma^2/2$). Hence, for further simplicity, we adopt the mean-field approximation $\mu^{(\rm{ex})}_{vdW} \approx \langle \varepsilon \rangle$.

The isothermal compressibility, $\kappa_T$, is given by 
\begin{eqnarray}
\kappa_T = \frac{ \langle V^2 \rangle - \langle V \rangle^2 } {k_{\rm B}T \langle V \rangle} = \frac{\sigma_V^2}{k_{\rm B}T \langle V \rangle} \, 
\end{eqnarray}
where $\sigma_V$ is the second central moment of the distribution of $V$ obtained in the simulations. If the sample of $V$ obtained from the simulation is independent, identically distributed (iid), then we know that $\sigma_V$ is $\chi^2$ distributed\cite{sivia} with the optimal value of $\sigma_V$ given by $\sigma_V = \sigma_0 \pm \sigma_0 / \sqrt{2(N-1)}$, where $\sigma_0$ is the sample standard deviation of $V$ and $N$ is the sample size.  However, the time-series trace of volumes in the simulation log is correlated. To this end, we compute the autocorrelation of $\delta V = V - \langle V\rangle$, and define the correlation length as the number of entries, $n$, in the log-file it takes for the normalized autocorrelation of $\delta V$ to fall below 0.05. 
(An alternative choice for the correlation length is the statistical inefficiency from the Friedberg-Cameron approach; this choice is slightly tighter, but it still leads to the same mean $\kappa_T$ and similar uncertainties.) We then sub-sample the time-series trace of V, such that the sampled $V$ are separated by the auto-correlation length.
For this sub-sample, we have $\sigma_V^2 \approx \sigma_0^2 \pm \sigma_0^2 \sqrt{2/(N-1)}$. (See also Ref.\ \citenum{smit:error}.) By shifting the time origin, we construct $n-1$ such sub-samples and compose the mean $\sigma_V^2$.  The error of the mean $\sigma_V^2$ is then obtained using variance propagation rules. 

\section{Results and Discussion}

\subsection{Bulk water}
\begin{figure}
\includegraphics[scale=0.95]{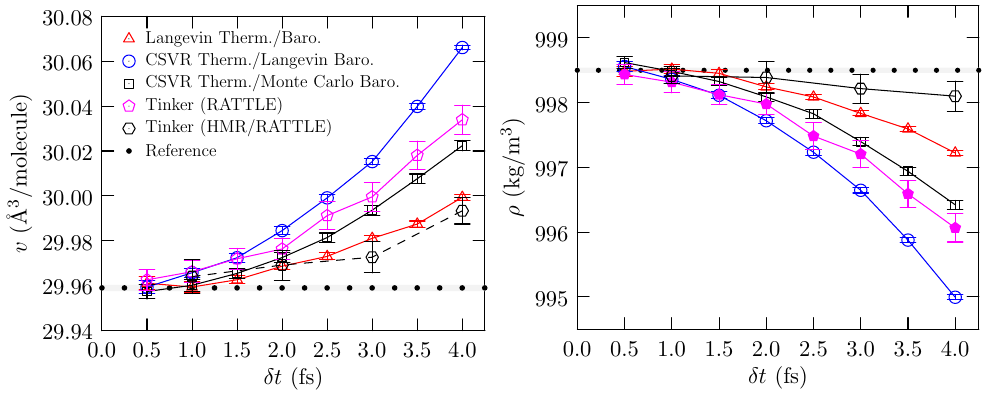}
\caption{Water partial molar volume $v$ (left panel) and mass density (right panel) versus $\delta t$ for simulations with thermostat set at 298.15~K and barostat set at 1 bar. The simulations with Tinker use the CSVR thermostat and the Monte Carlo barostat. The standard error of the mean is shown at the $2\sigma$ level. \underline{Reference}:  32768 waters, $\delta t = 0.25$~fs. Data for different volume sampling frequencies are averaged and standard errors appropriately propagated. The radius of the symbol ($\bullet$) is $2\sigma$ standard error of the mean.}
\label{fg:SPCEVol}
\end{figure}

Figure~\ref{fg:SPCEVol} shows the volume (density) versus $\delta t$ for different thermostat/barostat combinations. The horizontal line in the figure is the value obtained from averaging the four separate simulations for the large reference system.  We find that  by between 1250 steps for a volume sampling frequency of 80 and 2500 steps for a sampling frequency 200 the system
volume settles close to the  eventual mean value. This also helps confirm that the sample sizes for the reference calculation and the studies with the 4096 water system are rather conservative. 

Figure~\ref{fg:SPCEVol} shows that the volume increases with increasing $\delta t$, irrespective of the thermostat/barostat combination. This is consistent with our hypothesis based on observations in the earlier study that the translational temperature increases as $\delta t$ increases, and Davidchack's calculation of pressure versus $\delta t$ in the $NVT$ ensemble. 
Clearly, only for a small $\delta t$ --- for the conditions tested here $\delta t = 0.5$~fs ---  the volumes (densities) converge 
to a common value that is independent of the thermostat/barostat combination. Importantly, the volumes (densities) converge to the value obtained 
from the entirely separate reference simulation.  Lastly, hydrogen mass repartitioning (HMR) is an improvement in estimating the equilibrium volume (density), but the procedure is ineffective for $\delta t = 4$~fs. HMR can be defensible for  $\delta t \leq 2.0$~fs. 
 
As a further check, we sought to compare with predictions of density from an entirely stochastic simulation.  
Sanz et~al.\ \cite{sanz:prl04} have systematically explored the phase equilibrium of water for different water models using Monte Carlo calculations.  For the SPC/E model and for a nearly identical treatment of inter-molecular interactions --- they use a shorter cutoff of 8.5~{\AA} versus 9~{\AA} used by us --- they quote a density of 1000 kg/m$^3$ for 1~bar pressure and 300~K.  Using the experimental thermal expansion coefficient for water, we can infer that at 298.15~K, the density from the Monte Carlo procedure should be about 1000.4 kg/m$^3$. This value is an upper bound to the values obtained in the molecular dynamics simulations, with the least deviation of 0.2\% of 998.5~kg/m$^3$ obtained for $\delta = 0.5$~fs.  While this comparison is encouraging,  we must note some caveats. First, Sanz~et al.\  do not report the number of water molecules used in the simulations; we suspect\cite{vega:tip4p} it was considerably less than 4096, perhaps being as low as 360. As noted in our study on system size dependence of protein hydration \cite{asthagiri:jpcb20a},  a larger system better accommodates density fluctuations and this may explain part of the deviation. (Exploring the relevance of this issue for converged density predictions is left for a separate study.) Second, Sanz~et al.\  do not quote statistical uncertainties. We suspect\cite{vega:tip4p} it is about 1~kg/m$^3$,  in which case the agreement with our reference and $\delta t = 0.5$~fs results is
satisfactory within the quoted statistical uncertainties of the respective simulations. 

Consider next the change in the partial molar volume between $\delta t = 0.5$~fs and the more conventional $\delta t = 2.0$~fs. For simulations with the  CSVR thermostat and Monte Carlo barostat, the partial molar volume for $\delta t = 2.0$~fs is about 0.02~{\AA}$^3$ larger. Thus for a system with 10,000 water molecules, a system size that is nowadays rather common and likely on the smaller size-scale of simulation systems, the volume for the $\delta t = 2.0$~fs simulation will be larger by about 200~{\AA}$^3$ relative to that for the system simulated with $\delta t = 0.5$~fs. To put this deviation in perspective, we note the following example. The volume change upon folding of the 149 residue Staphylococcal Nuclease protein at 21$^\circ$C, close to the temperature studied here, is found to be about $70$ ml/mol or about $116$~{\AA}$^3$/molecule (of S.\ Nuclease) \cite{panick:sn1999,paliwal:sn04}. This deviation is already smaller than the error
in overall system volume induced by too large of a $\delta t$.  Volume change upon folding/unfolding for similarly sized or smaller proteins will be comparable or smaller.  Thus the impact of the artifacts due to too large a $\delta t$ will be proportionally greater in assessing both the thermodynamics and the kinetics of the folding/unfolding transition in computer simulations. 

\begin{figure}
\includegraphics[scale=0.95]{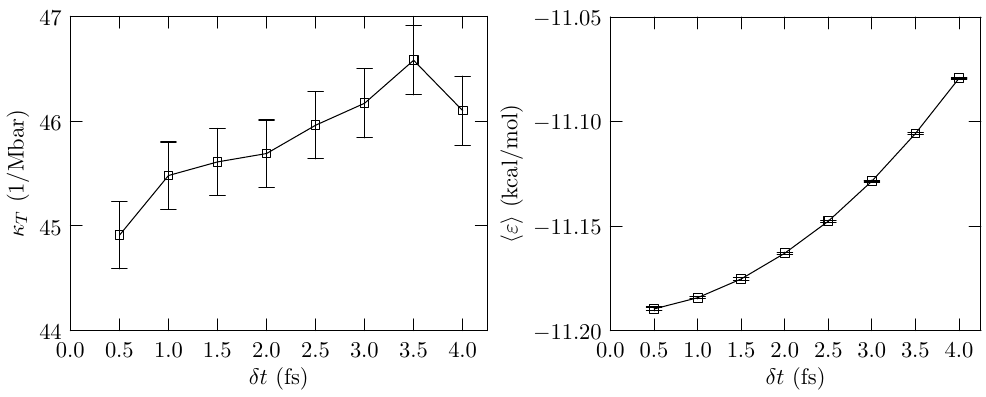}
\caption{\underline{Left panel}: Calculated compressibility versus $\delta t$.  The uncertainty is shown at the $1\sigma$ level. \underline{Right panel}: The binding energy of a water molecule averaged over all frames.  The standard error of the mean is shown at the $2\sigma$ level. The $NpT$ calculations are performed with the CSVR thermostat and the Monte Carlo barostat within the NAMD program.}
\label{fg:chi}
\end{figure}

Since the volume depends on $\delta t$, it begs the question whether the fluctuation in the volume under 
$NpT$ conditions also depends on $\delta t$. Figure~\ref{fg:chi} shows the behavior of the estimated compressibility. For this analysis, we only consider CSVR thermostat/Monte Carlo barostat; 
in contrast to the Langevin thermostat the CSVR thermostat is less intrusive in the dynamics and affects translation and rotation symmetrically. (With the CSVR thermostat the average of the translation and rotation temperatures equals the thermostat set-point temperature, unlike what we find for the Langevin thermostat\cite{asthagiri:jctc2024a}.) 

In Figure~\ref{fg:chi} (left panel), the estimated statistical uncertainties are large, as expected given the overall length of the simulation after accounting for correlations; however, the trend is unmistakable: the compressibility tends to decrease with decreasing $\delta t$. The behavior of the compressibility with $\delta t$ is also consistent with the behavior of the binding energy with $\delta t$ (right panel):  as cohesion increases one expects the fluid matrix to become stiffer and the compressibility to decrease. 

From $NVT$ simulations, we can compute the dielectric constant using the fluctuation of the dipole moment of the simulation cell. This quantity also shows a dependence on $\delta t$ (data not shown), but the statistical uncertainties are large. However, we can infer the dependence of the dielectric constant on $\delta t$ from the purely electrostatic contribution to the free energy of hydration. We have tested this for di- and tri-valent ions (data not shown). Below we show the same behavior for the more interesting case of the hydration of a protein. 

\subsection{Aqueous BBA} 
\begin{figure}
\includegraphics[scale=0.95]{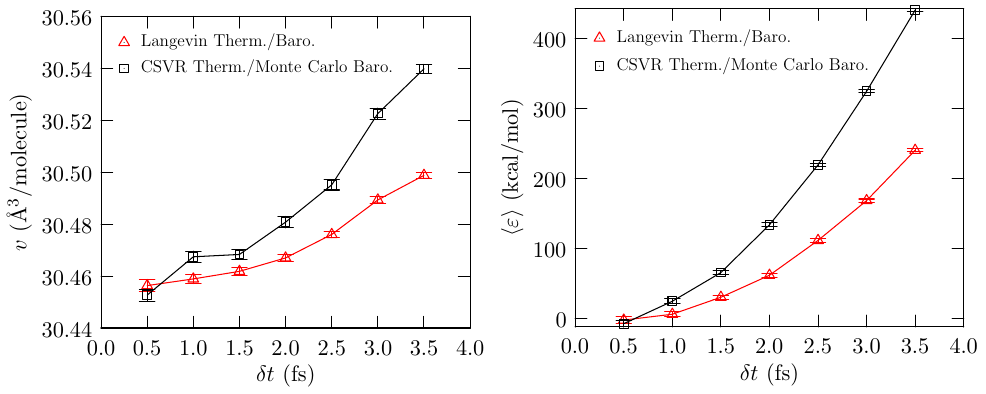}
\caption{\underline{Left panel}: Dependence of partial molar volume $v$ of water in the presence of one molecule of BBA. \underline{Right panel}: Total mean potential energy of the system relative to the value for $\delta t = 0.5$~fs. Thermostat is set at 298.15~K and barostat is set at 1 bar. The standard error of the mean is shown at the $2\sigma$ level.}
\label{fg:BBAVol}
\end{figure}

 Figure~\ref{fg:BBAVol} (left panel) shows the partial molar volume of water versus $\delta t$ for the aqueous protein system. Just as we found for bulk water, the simulation volume converges to a value independent of the thermostat/barostat combination only for a small $\delta t$. For the conditions tested here, $\delta t = 0.5$~fs ensures convergence of volume. It is heartening that this value of $\delta t$ 
 also ensure the convergence of the system potential energy to a common value. 
 
 Examining the dependence of the radius of gyration of the protein ($R_g$) on $\delta t$ in carefully constructed replica exchange simulations proved to be inconclusive in revealing systematic trends (data not shown). This can be due to confounding roles of flexibility, distribution of the counterion (Cl$^{-}$), and the potential for artifacts introduced by violation of canonical sampling\cite{hummer:replicathermostat2009}. To better isolate the dependence of protein-solvent interactions  on $\delta t$, we consider a protein in a fixed conformation in a water bath and without the counterion. 
 
 Table~\ref{tb:charging} shows the average equilibrium box length obtained from $NpT$ calculations for this system. These values 
 are used in the $NVT$ ensemble calculations of $\mu^{({\rm ex})}_{elec}$  and $\mu^{({\rm ex})}_{vdW}$.
 \newcommand{\TT}[2]{\ensuremath{{#1}\pm{#2}}}
 \setlength{\tabcolsep}{10pt}
 \ctable[
	mincapwidth=\textwidth,
	caption={Equilibrated volumes with standard error of the mean at $1\sigma$ level and the 
	box length of the simulation with the protein in a fixed conformation. The Wigner self-interaction correction\cite{Hummer:ions1996,Hummer:ions1998} is $\mu^{({\rm ex})}_{self} = (1/2) \xi\sum_i q_i^2$, where $q_i$ is the partial charge of atom $i$ of the protein and $\xi = -2.837297 / L$. We report this correction relative to the $\delta t = 0.5$~fs case.},
	label=tb:charging,
	pos=h!,
	captionskip=-1.ex
	]	
{c c c c}
{}
{
\FL
$\delta t$ (fs)     &  V  ({\AA}$^3$)     &     $L$ ({\AA})       &  $\mu^{({\rm ex})}_{self}$ (kcal/mol)        \ML
0.5                     &   \TT{200074}{13}  &        58.488         & 0.0                                     \NN[-1ex]
1.0                     &   \TT{200109}{9}   &          58.491         & 0.0                                  \NN[-1ex]
2.0                    &    \TT{200289}{7}   &           58.508         & 0.2                                  \NN[-1ex]
3.0                    &    \TT{ 200571}{6}   &         58.536          & 0.4                                    \NN[-1ex]
4.0                    &    \TT{200978}{6}    &         58.576          & 0.7                                    \LL
}
For the usual choice of $\delta t = 2.0$~fs relative to $\delta t = 0.5$~fs we can ignore the Wigner self-interaction correction for this system. Also, since the box is relatively large compared to the size of the protein, we ignore the finite-size correction to electrostatic interactions. 

\begin{figure}[h!]
\includegraphics[scale=0.95]{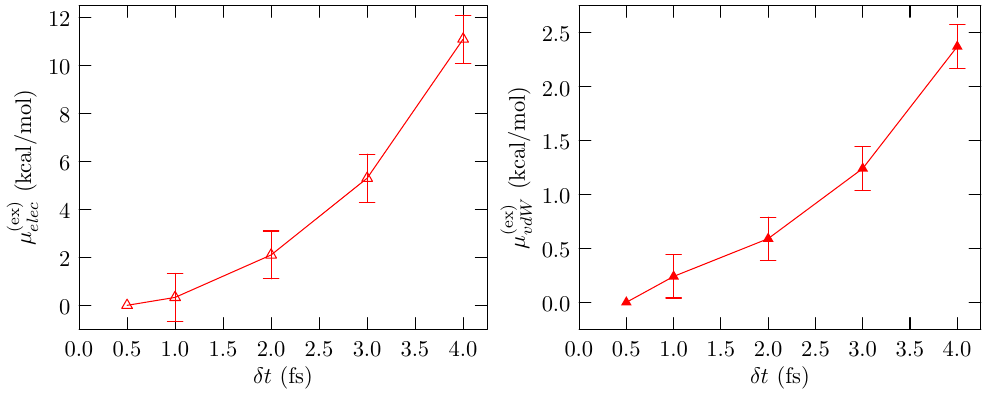}
\caption{\underline{Left panel}: Dependence of electrostatic contribution to the free energy of hydration on $\delta t$.  The Wigner self-interaction correction is {not} included.  \underline{Right panel}: $\delta t$ dependence of mean-field van der Waals contribution to the hydration free energy.  Throughout the numbers are shown relative to the mean value at $\delta t = 0.5$~fs. The standard error of the mean is shown at the $1\sigma$ level.}
\label{fg:BBAmu}
\end{figure}

Figure~\ref{fg:BBAmu} shows $\mu^{({\rm ex})}_{elec}$  and $\mu^{({\rm ex})}_{vdW}$ of BBA.  
First, consider the electrostatic contribution. We find that the linear response result already gives $\approx95$\% of the electrostatic contribution (data not shown), emphasizing that the three point rule is adequate in describing the charging free energy.  Second, the behavior across $\delta t$ shows that 
$\mu^{({\rm ex})}_{elec}$ is sensitive to $\delta t$, even without including the Wigner self-interaction correction (Table~\ref{tb:charging}) which amplify the trend even more. Finally, 
the vdW contribution obtained using the mean-field approximation $\mu^{(\rm{ex})}_{vdW} \approx \langle \varepsilon \rangle$ is also sensitive to the step-size, although, as expected, not as strongly as $\mu^{({\rm ex})}_{elec}$. 

Shaw and coworkers\cite{deshaw:sc2011} had studied the folding/unfolding transition of the BBA protein in very long computer simulations. At 325~K, they have calculated a folding free energy of about 0.7~kcal/mol. (That study used a time-step of 2.5~fs.) Wu and Shea have estimated a similar value at 323~K \cite{shea:bba2010}. Assuming the folding free energy of the BBA protein is in a similar range at 298.15 K as well, we can see that the error introduced by a larger step-size in the calculation of the hydration free energy is already considerably larger than the reported folding free energy. How these errors behave in unfolded$\leftrightarrow$folded transition kinetics and thermodynamics is  left for future studies.

The earlier work\cite{asthagiri:jctc2024a} and the present  show that in a molecular dynamics approach to sampling equilibrium ensembles, it is important to capture the relevant relaxation dynamics with fidelity. However, in molecular dynamics sampling --- and, for simplicity, we consider the $NVE$ ensemble --- the relevant metrics are conserved quantities, the foremost being the energy of the system; fidelity of capturing relaxation dynamics does not  appear explicitly. 
In this regard, note that in discrete Hamiltonian dynamics with time-reversible algorithms, the sampling is from a shadow Hamiltonian, $\tilde{\mathcal{H}}(\delta t) = \mathcal{H} + G(\delta t^2)$, where $\mathcal{H}$ is the physical Hamiltonian \cite{sanz-serna:Acta92,toxvaerd:pre94,gans:pre2000,hairer:Acta03}.  (The general form of $G$ is unknown but is usually investigated by means of a series expansion\cite{sanz-serna:book}.) As a consequence, and within time-reversible algorithms, $\mathcal{H}$ is not conserved but oscillates about a mean value, with the size of oscillations proportional to $\delta t^2$ \cite{hairer:Acta03}. This discretization error, or what can also be thought of as an exchange of ``shadow" work with the system\cite{sivak:prx13}, 
is what impacts the relaxation.  Thus, we see that an algorithm can be globally stable with a long $\delta t$, but that stability is \emph{not sufficient} to ensure relaxation processes are correctly captured. 

A further consequence of discrete Hamiltonian dynamics is that the particle velocity $v \neq p / m$, where $p$ is the momentum obtained using the ``Velocity"-Verlet equation\cite{gans:pre2000} and $m$ is its mass. Better estimates of $v$ can be obtained 
and the temperature defined using the  generalized equipartition theorem\cite{shaw:jctc2010,tolman:equip1918}. This is an important point to consider. However, the key insight from the earlier work and the present one is the 
need to obey the fluctuation-dissipation theorem, i.e.\ the need to capture the temporal evolution of fluctuations\cite{pathria} or equivalently the underlying relaxation dynamics with fidelity in a MD sampling of ensembles. 

For a molecular dynamics sampling of equilibrium ensembles, we suggest it would be prudent to study relevant velocity autocorrelations\cite{asthagiri:jctc2024a}, or if that is too tedious, estimate quantities such as mean energy or volume for several different time-steps to ensure that time-step artifacts are under control. Such checks would be especially prudent
before undertaking large-scale simulation campaigns. 

\section{Conclusions}
It has been about 47 years since Ryckaert, Ciccotti, and Berendsen introduced the SHAKE algorithm that allowed the efficient description 
of molecules such as water as a rigid object in computer simulations. A guiding idea in this development  was the innocuous assumption that fast internal vibrations in the molecule are decoupled from translational and vibrational modes, and hence after rigidifying the molecule, it is permissible to take a longer time-step to integrate the equations of motion.  This assumption is not valid for water or solutes dissolved in water. It is humbling, and also encouraging, that the early pioneers in the molecular dynamics simulation of water, Rahman and Stillinger, suggested a time step of $\delta t = 0.4$~fs for their  model of water, close to the value of 0.5~fs we find for both SPC/E and TIP3P models. 

A long time-step for integrating the equations of motion leads to the breakdown of equipartition in describing water. The rapid angular motion of water relative to the translation of the center-of-mass ensures that the rotational relaxation occurs at a far shorter time scale than the translation relaxation. This physics  imposes a fundamental limit on how long a time-step one can use for correctly capturing the temporal evolution of fluctuations, as is required for adhering to the fluctuation-dissipation relation. 

The breakdown of equipartition leads to the translation modes being hotter than the rotational modes, a deviation that is amplified as the time-step is increased. As a consequence, at constant pressure, the volume is larger for longer time-step. For a protein dissolved in water, the deviation in the partial molar volume 
 will necessarily introduce uncontrolled $p-V$ errors in the folding free energy landscape. 

The $\delta t$ artifact also impacts the interaction between the protein and solvent. For the BBA protein, the error introduced in the  electrostatic plus vdW contribution to the hydration free energy can easily exceed the folding free energy. Thus, too long a time-step introduces errors in the interaction contribution, further obfuscating the resolution of the free energy landscape.

\section{Acknowledgements}
We thank Margaret Whitt (NGSI Intern at ORNL) for exploratory studies on protein BBA. We thank Jindal Shah for helpful pointers on Monte Carlo simulations with the Cassandra code.  We thank Thiago Pinheiro dos Santos for numerous helpful discussions. We thank David Rogers for a critical reading of the manuscript and helpful comments.  This research used resources of the Oak Ridge Leadership Computing Facility at the Oak Ridge National Laboratory, which is supported by the Office of Science of the U.S. Department of Energy under Contract No. DE-AC05-00OR22725.
 
 \clearpage

\providecommand{\latin}[1]{#1}
\makeatletter
\providecommand{\doi}
  {\begingroup\let\do\@makeother\dospecials
  \catcode`\{=1 \catcode`\}=2 \doi@aux}
\providecommand{\doi@aux}[1]{\endgroup\texttt{#1}}
\makeatother
\providecommand*\mcitethebibliography{\thebibliography}
\csname @ifundefined\endcsname{endmcitethebibliography}
  {\let\endmcitethebibliography\endthebibliography}{}

 \end{document}